\documentclass[aps,prl,twocolumn,showpacs]{revtex4}
\usepackage[T1]{fontenc}
\usepackage[latin1]{inputenc}
\usepackage{graphics}
\usepackage{amsmath}

\makeatletter

\providecommand{\LyX}{L\kern-.1667em\lower.25em\hbox{Y}\kern-.125emX\@}

\makeatother

\begin{document}

\title{$q$-breathers in finite
two- and three-dimensional nonlinear acoustic lattices}
\author{M.~V.~Ivanchenko$^1$, O.~I.~Kanakov$^1$, K.~G.~Mishagin$^1$ and S. Flach$^2$}

\affiliation{$^1$ Department of Radiophysics, Nizhny Novgorod
University, Gagarin Avenue, 23, 603950 Nizhny Novgorod, Russia \\
$^2$ Max-Planck-Institut f\"ur Physik komplexer Systeme,
N\"othnitzer Str. 38, D-01187 Dresden, Germany }

\begin{abstract}
Nonlinear interaction between normal modes dramatically affects
energy equipartition, heat conduction and other fundamental
processes in extended systems. In their celebrated experiment Fermi,
Pasta and Ulam (FPU, 1955) observed that in simple one-dimensional
nonlinear atomic chains the energy must not always be equally
shared among the modes. Recently, it was shown that exact and
stable time-periodic orbits, coined $q$-breathers (QBs), localize
the mode energy in normal mode space in an exponential way, and
account for many aspects of the FPU problem. Here we take the
problem into more physically important cases of two- and three-dimensional
acoustic lattices to find existence and principally different
features of QBs. By use of perturbation theory and numerical
calculations we obtain that the localization and stability of QBs
is enhanced with increasing system size in higher lattice
dimensions opposite to their one-dimensional analogues.
\end{abstract}


\maketitle

Nonlinearity induced interaction between normal modes of
extended systems is crucial for many fundamental dynamical and
statistical phenomena like thermalization, thermal expansion of
solids or turbulence in liquids. It is also important in many
artificial systems where one aims at controlling the energy flow
among the normal modes, preventing or efficiently channeling
energy pumping due to resonances with external perturbations 
etc. Among the many accumulated results in this area a seminal one
is due to Fermi, Pasta and Ulam (FPU) who reported on the absence
of thermalization of chains of atoms connected by weakly nonlinear
springs \cite{fpu}. They observed that the energy of an initially
excited normal mode with frequency $\omega_q$ and wave number $q$
did not spread over all other normal modes, staying almost
completely locked within a few neighbouring modes in the normal
mode space \cite{Ford,chaosfpu}. Longer waiting times yielded
another puzzle of energy reccurrence to the originally excited
mode. Many efforts to explain the FPU paradox resulted in an
extraordinary progress in this field: the observation of solitons
\cite{Zabusky}, size-dependent stochasticity thresholds
\cite{Izrailev_Chirikov}, nonlinear resonances \cite{Chirikov},
KAM tori, Arnold diffusion, and many other issues
\cite{deLuca,Shepel,italian,kantz,lcmcsmpegdc97}. The efforts to
carry all these concepts into two-dimensional lattices have been
also reported \cite{2dim}. However, according to Ford \cite{Ford},
despite the richness of new topics, the original FPU problem was
still waiting to be understood.

Recently it was shown that the major ingredients of the FPU
problem can be addressed within the promising concept of
$q$-breathers (QBs), which are exact time-periodic solutions in
the nonlinear FPU chain \cite{we}. These solutions are
exponentially localized in the $q$-space of normal modes and
preserve stability for small enough nonlinearity. They continue
from their trivial counterparts for zero nonlinearity at finite
energy. In that limit they correspond to one mode with the seed
wave number $q_0$ being excited, and all the other modes being at
rest. The stability threshold of QB solutions coincides with the
weak chaos threshold in \cite{deLuca}. Persistence of exact stable
QB modes surrounded by almost quasiperiodic trajectories explains
the origin of FPU recurrences and the absence of energy
equipartition. The scaling of the localization exponents with the
relevant control parameters relates to results on higher order
nonlinear resonance overlaps \cite{Shepel}. But perhaps the most
important result was, that it needs only one ingredient to obtain
QBs in FPU chains: a discrete nonequidistant frequency spectrum of
normal modes, as induced by a finite system. Thus the door is
opened to apply the concept of QBs to a variety of nonlinear
finite systems - truly common objects in nature and applications.
One of such challenges is an extension of the notion of QBs into
more physically realistic two- and three-dimensional FPU-type
systems.

In this Letter we report on the existence and remarkable
properties of $q$-breathers in
 finite two-and three-dimensional nonlinear acoustic lattices.
For fixed energy, nonlinearity coefficient and seed wave number
the QB localization length stays finite in the 2d case  and tends
to zero in the 3d case with increasing size of the system. For
both 2d and 3d cases the nonlinearity coefficient at the QB
stability threshold increases in the same limit. This comprises a
crucial difference from the 1d case \cite{we} where QBs delocalize
and the nonlinearity threshold of the instability tends to zero in
the limit of large chains.

We consider quadratic and cubic lattices
of $N^d$ ($d=2$ and $3$ respectively) equal masses coupled by
nonlinear springs with the Hamiltonian
\begin{equation}
\label{eq1}
\begin{aligned}
& H=\frac{1}{2}\sum\limits_{\boldsymbol{n}}(p_{\boldsymbol{n}}^2
+\sum\limits_{\boldsymbol{m}\in D(\boldsymbol{n})}[
\frac{1}{2}(x_{\boldsymbol{m}}-x_{\boldsymbol{n}})^2+\frac{\beta}{4}(x_{\boldsymbol{m}}-x_{\boldsymbol{n}})^4])
\end{aligned}
\end{equation}
where $x_{\boldsymbol{n}}(t)$ is the displacement of the
$\boldsymbol{n}=(n_1,\ldots,n_d)$-th particle from its original
position, $p_{\boldsymbol{n}}(t)$ its momentum,
$D(\boldsymbol{n})$ is the set of its nearest neighbors, and fixed (zero)
boundary conditions are taken: $x_{\boldsymbol{n}}=0$ if $n_l=0$ or
$n_l=N+1$ for any of the components of $\boldsymbol{n}$.

 A canonical transformation
 \begin{equation}
 \label{eq1.5}
x_{\boldsymbol{n}}(t)=\left(\frac{2}{N+1}\right)^{d/2}\sum\limits_{q_{1},\ldots,
q_{d}=1}^N Q_{\boldsymbol{q}}(t)\prod\limits_{i=1}^d
\sin{\left(\frac{\pi q_i n_i}{N+1}\right)}
\end{equation}
 takes into the
reciprocal wave number space with $N^d$ normal mode coordinates
$Q_{\boldsymbol{q}}(t)\equiv Q_{q_1,\ldots,q_d}(t)$.
The normal mode space is spanned by $\boldsymbol{q}$ and represents
a d-dimensional lattice similar to the situation in real space.
The equations
of motion then read
\begin{equation}
\label{eq2}
 \Ddot{Q}_{\boldsymbol{q}}+\Omega_{\boldsymbol{q}}^2
 Q_{\boldsymbol{q}}=-\frac{16\beta}{(2N+2)^d}\sum\limits_{\boldsymbol{p},\boldsymbol{r},\boldsymbol{s}}
 C_{\boldsymbol{q},\boldsymbol{p},\boldsymbol{r},\boldsymbol{s}}Q_{\boldsymbol{p}}Q_{\boldsymbol{r}}
 Q_{\boldsymbol{s}}\;.
\end{equation}
Here
$\Omega_{\boldsymbol{q}}^2=4\sum\limits_{i=1}^d\omega_{q_i}^2$ are
the squared normal mode frequencies with $\omega_{q_{i}}=\sin{(\pi
q_{i}/2(N+1))}$.
Note, that all linear modes but the diagonal ones $Q_{q_1=\ldots=q_d}$
are at least $d$-fold degenerate with respect to their frequencies.
The coupling coefficients
$C_{\boldsymbol{q},\boldsymbol{p},\boldsymbol{r},\boldsymbol{s}}$
induce a selective interaction between distant
modes in the normal mode space similar to the 1d case.

 For small amplitude excitations the nonlinear terms in
the equations of motion can be neglected, and according to
(\ref{eq2}) the $q$-oscillators get decoupled, each conserving its
harmonic energy
$E_{\boldsymbol{q}}=\frac{1}{2}\left(\dot{Q}_{\boldsymbol{q}}^2+\Omega_{\boldsymbol{q}}^2
Q_{\boldsymbol{q}}^2\right)$ in time. Especially, we may consider
the excitation of only one of the $q$-oscillators, i.e.
$E_{\boldsymbol{q}} \neq 0$ for $\boldsymbol{q}\equiv
\boldsymbol{q}_0$ only. Such excitations are trivial time-periodic
and $q$-localized solutions (QBs) for $\beta=0$.

For the 1d chain, such periodic orbits can be continued into the
nonlinear case at fixed total energy \cite{we} because the
nonresonance condition $n \Omega_{\boldsymbol{q}_0} \neq
\Omega_{\boldsymbol{q} \neq \boldsymbol{q}_0}$ (here $n$ is an
integer) holds for any finite size \cite{Tiziano}. This argument
can be used straightforwardly for $d=2,3$ and large seed wave
numbers $\boldsymbol{q}$ on the main diagonal such that
$2\Omega_{\boldsymbol{q}} > 4$. For all other seed wave numbers on
the main diagonal we checked that the nonresonance condition holds
as well. For seed wave numbers off the main diagonal the above
mentioned $d$-fold degeneracies could be lifted by considering
anisotropic lattices. In fact these
degeneracies are lifted by the nonintegrability of the nonlinear
lattice (\ref{eq2}), and a discrete set of periodic orbits will
remain. QBs are successfully observed in numerical experiments in
the presence of the degeneracy and we did not find substantial
difficulties in computing them.

In the following we compute QBs as well as their Floquet spectrum
numerically using the same algorithms as for the 1d atomic chain \cite{we},
and compare the results with analytical predictions, derived by
means of asymptotic expansions.

\begin{figure}[t]
{\centering
\resizebox*{0.90\columnwidth}{!}{\includegraphics{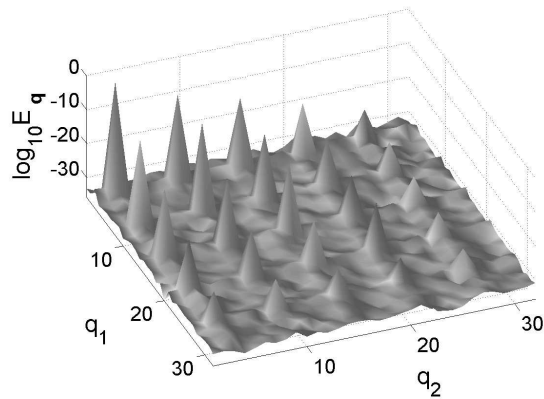}}}
{\caption{A QB for $d=2$ with $\boldsymbol{q}_0=(3,3)$ for
$\beta=0.5, E_{tot}=1.5$, frequency $\hat{\Omega}\approx0.403,
N=32$.}\label{fig1}}
\end{figure}

\begin{figure}[t]
{\centering
\resizebox*{0.90\columnwidth}{!}{\includegraphics{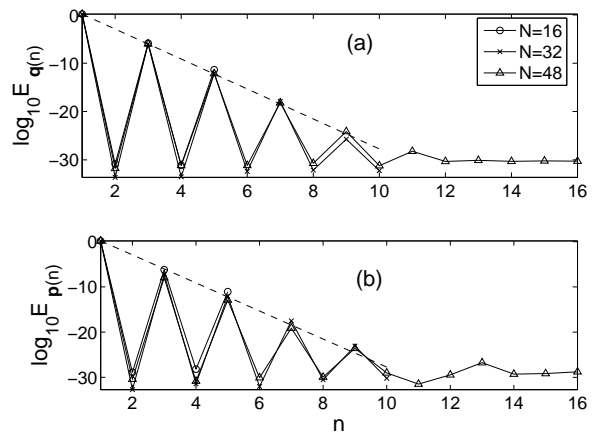}}}
{\caption{The normal mode energies along (a) the diagonal
$\boldsymbol{q}(n)=n\boldsymbol{q}_0$ and (b) the side
direction $\boldsymbol{p}(n)=(q_{0,1},nq_{0,2})$ for a QB
with $\boldsymbol{q}_0=(3,3), \beta=0.5, E_{tot}=1.5, N=16, 32,
48$ and the analytical estimate (\ref{eq4}) (dashed lines).}
\label{fig2}}
\end{figure}

First, we turn to the case $d=2$. We obtain
various symmetric (with $(q_0)_1=(q_0)_2$, Fig.\ref{fig1},\ref{fig2})
and asymmetric (with
$(q_0)_1\neq(q_0)_2$) QBs in the lower frequency range, which are
exponentially localized in $q$-space.
Note,
that due to the parity symmetry of the model
(Eq.(\ref{eq1}) is invariant under $x_{\boldsymbol{n}} \rightarrow
-x_{\boldsymbol{n}}$ for all $\boldsymbol{n}$) only the modes with
odd components $(q_1,q_2)$ are excited by the $\boldsymbol{q}_0=(3,3)$ mode.
In contrast to $d=1$, the decay of the energy distribution
(especially along the diagonal
$\boldsymbol{q}(n)=(2n-1)\boldsymbol{q}_0$) remains almost
constant with increase of the lattice size (Fig.\ref{fig2}).

The abovementioned degeneracy of the frequency spectrum supports
the existence of multi-mode QBs, namely those, which have two (or
more) excited seed modes. Indeed, we continued multi-mode periodic
solutions of the linear lattice $E_{\boldsymbol{q}} \neq 0$ for
$\boldsymbol{q}\in S(\boldsymbol{q}_0)=\{\boldsymbol{q}:
\Omega_{\boldsymbol{q}}=\Omega_{\boldsymbol{q}_0}\}$ into the
nonlinear regime. For example, the set
$\boldsymbol{q}_0=(2,3),\boldsymbol{q}_0^*=(3,2)$ allows for two
(asymmetric single-mode) QB solutions with the energy mainly
concentrated in one of the two seed modes and two symmetric
multi-mode QB solutions with the same energy in each of the seed
modes, and oscillations being in- or out-of-phase.

By an asymptotic expansion of the solution to (\ref{eq2}) in
powers of the small parameter $\sigma=\beta/(N+1)^2$ (in analogy to
\cite{we}) we estimate
the shape of a QB solution $\hat{Q}_{\boldsymbol{q}}(t)$ with
a low-frequency seed mode number $\boldsymbol{q}_0$. The energies of the
modes on the diagonal of the QB $\boldsymbol{q}_0$,
$3\boldsymbol{q}_0$,\dots,$(2n+1)\boldsymbol{q}_0$,\dots$\ll(N,N)$
read
\begin{equation}
\label{eq4}
E_{(2n+1)\boldsymbol{q}_0}=\lambda_{d}^{2n}E_{\boldsymbol{q}_0}\;,\;
\lambda_{d}=\frac{3\beta
E_{\boldsymbol{q}_0}N^{2-d}}{2^{2+d}\pi^2|\boldsymbol{q}_0|^2}\;\;.
\end{equation}
Dashed lines in Fig.\ref{fig2}(a) are obtained using (\ref{eq4})
and show very good agreement with the numerical results. The
energy distribution between other modes, involved in the QB is
more complicated, but the decay along the diagonal
(\ref{eq4}) gives a good estimate for it (Fig.\ref{fig2}(b)).
Note, that the localization along the diagonal is the weakest, at
least for large $N$. The shape of the QB in the $q$-space then
possesses the following properties: (i) the localization remains
constant when the lattice size tends to infinity with all other
parameters fixed, in contrast to the 1d case where QBs delocalize as
$\lambda\propto(N+1)$, (ii) in the limit of constant {\it energy
density} $\varepsilon=E_{\boldsymbol{q}_0}/(N+1)^2$ and {\it wave
vector} of the QB $\boldsymbol{\kappa}_0=\boldsymbol{q}_0/(N+1)$
the localization remains constant for
$N\rightarrow\infty$ (similar to the 1d case), (iii) for
smaller $\beta, E_{\boldsymbol{q}_0}$
and larger $\boldsymbol{q}_0$ QBs compactify. These results
are a consequence of a more fundamental scaling property
of finite to infinite systems.

\begin{figure}[t]
{\centering
\resizebox*{0.90\columnwidth}{!}{\includegraphics{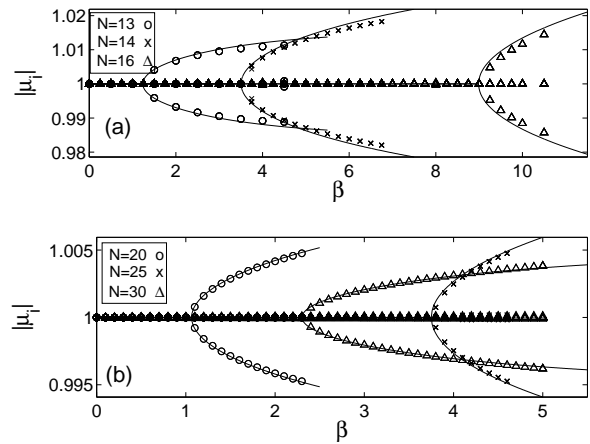}}}
{\caption{Absolute values of Floquet eigenvalues (symbols) and analytical
curves (solid lines) for QBs with
$E_{tot}=1.5$ and (a) $\boldsymbol{q}_0=(3,3), N=13, 14, 16$ (the
instabilities correspond to $\boldsymbol{k}=(1,2)$ and
$\boldsymbol{k}=(2,1)$) and (b) $\boldsymbol{q}_0=(2,3), N=20, 25,
30$ (the instabilities correspond to $\boldsymbol{k}=(1,1)$ for
$N=20, 25$ and $\boldsymbol{k}=(1,2)$ for $N=30$).} \label{fig4}}
\end{figure}

We analyze the stability of the obtained periodic orbits with
standard methods of linearizing the phase space flow around the
solutions and computing the eigenvalues and eigenvectors of the
corresponding symplectic Floquet matrix \cite{we}. The orbits are
stable if all complex eigenvalues lie on the unit circle. The
absolute values of the Floquet eigenvalues of QBs for symmetric
$\boldsymbol{q}_0=(3,3)$ and asymmetric $\boldsymbol{q}_0=(2,3)$
are plotted versus $\beta$ for different system sizes $N$  in
Fig.\ref{fig4}. Similar to the 1d case, QBs are stable for
sufficiently weak nonlinearities. When $\beta$ exceeds a certain
threshold $\beta^*$, some eigenvalues get absolute values larger
than unity (and some of them less than unity) and the QB becomes
unstable. In remarkable contrast to the 1d case, $\beta^*$ rapidly
increases with the size of the system Fig.\ref{fig4}(a). For a
series of computationally accessible large lattices $N=20, 30, 40$
(not plotted in Fig.\ref{fig4}(a)) we found the QB with
$\boldsymbol{q}_0=(3,3)$ to be stable at least up to $\beta=10.0$.
For insufficiently large $N$ the dependence $\beta^*(N)$ may
become non-monotonous (Fig.\ref{fig4}(b)). It may be quite
sensitive to the chosen seed wave number $\boldsymbol{q}_0$,
compare Figs.\ref{fig4}(a) and (b).

The observed instabilities can be traced analytically, similar to
the 1d case. Using standard secular perturbation techniques we
approximate the frequency of the QB solution  as $ \hat{\Omega} =
\Omega_{\boldsymbol{q}_0}(1+9h\rho)+O(h^2)$, where $h=3\beta
E_{\boldsymbol{q}_0}/(N+1)^2$ is a small parameter and
$\rho=(w_{(q_0)_1}^2+w_{(q_0)_2}^2)/\Omega_{\boldsymbol{q}_0}^4$.
Linearizing the equations of motion (\ref{eq2}) around a QB
solution
$Q_{\boldsymbol{q}}=\hat{Q}_{\boldsymbol{q}}(t)+\xi_{\boldsymbol{q}}$
we obtain
 \begin{equation}
  \label{eq4a}
 \Ddot{\xi}_{\boldsymbol{q}}+\Omega_{\boldsymbol{q}}^2
 \xi_{\boldsymbol{q}}=-4h(1+\cos{2\hat{\Omega}t})\sum\limits_{\boldsymbol{p}}
 C_{\boldsymbol{q},\boldsymbol{q}_0,\boldsymbol{q}_0,\boldsymbol{p}}/
 \Omega_{\boldsymbol{q}_0}^2\xi_{\boldsymbol{p}}+O(h^2)
\end{equation}
The strongest instability, caused by primary parametric resonance in
\eqref{eq4a}, comes from pairs of resonant modes
$\boldsymbol{q}+\boldsymbol{p}=2\boldsymbol{q}_0$ with a nonzero
vector
$\boldsymbol{k}=\boldsymbol{q}-\boldsymbol{q}_0=
\boldsymbol{q_0}-\boldsymbol{p}$.
The bifurcation point and the absolute values of the Floquet
multipliers involved in the resonance in its vicinity are
represented by a complex expression, which demonstrates good
agreement with the numerical results (Fig.\ref{fig4}).
We assume $|k_{1,2}|<<(q_0)_{1,2}<<N$ and
approximate
\begin{equation}
\label{eq5}
\begin{aligned}
&3\beta^*
E_{\boldsymbol{q}_0}/(N+1)^2=8(\Omega_{\boldsymbol{q}}+\Omega_{\boldsymbol{p}}-
2\Omega_{\boldsymbol{q}_0})/\Omega_{\boldsymbol{q}_0}\\
 & \approx 8\left(\frac{(q_0)_2k_1-(q_0)_1k_2}{\boldsymbol{q}_0^2}\right)^2
\end{aligned}
\end{equation}
This estimate explains the following basic features of the
observed instability. The instability of the type
$\boldsymbol{k}$ that minimizes $|(q_0)_2k_1-(q_0)_1k_2|$ is the
first to occur. It results in a non-monotonous and discontinuous
dependence of $\beta^*(\boldsymbol{q}_0,N)$ for small
lattices. It monotonously increases with $N$ while
$\boldsymbol{k}$ is constant (Fig.\ref{fig4}(a)); when
$\boldsymbol{k}$ changes $\beta^*$ changes as well (Fig.\ref{fig4}(b)).
Besides, the instability threshold scales as
$\beta^*\propto\boldsymbol{q}_0^{-2}$
and as $\beta^*\propto N^2$.

The 2d
lattice supports also the existence of QBs with $\boldsymbol{q}_0$
located in the intermediate
and high-frequency parts of the normal mode spectrum.

Let us turn to the case $d=3$. We again compute QB solutions
as in the 2d case. A similar analysis shows, that for the
low-frequency seed mode $\boldsymbol{q}_0$ the decay of the normal
mode energies along the leading direction
$3\boldsymbol{q}_0$,\dots,$(2n+1)\boldsymbol{q}_0$,\dots$\ll(N,N,N)$
is given by (\ref{eq4}) with $d=3$, and fits well the shape of the
numerically computed QBs with $\boldsymbol{q}_0=(3,3,3)$
(Fig.\ref{fig6}). In contrast to the 2d case, the localization
length even decreases with increasing lattice size
$\lambda_{3d}\propto(N+1)^{-1}$. For constant {\it energy density}
$\varepsilon=E/(N+1)^3$ and {\it wave vector} of the QB
$\boldsymbol{\kappa}_0$, the degree of localization is independent
of the lattice size, as it is for the 1d and 2d cases. Despite
having difficulties when calculating the stability of QBs in
sufficiently large 3d lattices due to limited machine performance,
the analytical treatment has been done similar to lower
lattice dimensions. We find that $\beta^*\propto(N+1)^3$ for
constant energy of the lattice and does not increase for fixed
$\varepsilon$ and $\boldsymbol{q}_0$. We also predict its
sensitive dependence upon the choice of the seed mode number of
the QB and the size of the lattice when being small.

The main reason for the crucial improvement of
localization and stability of QBs in higher dimensions relies on the fact
that the effective strength of nonlinear intermode coupling decreases
exponentially due to the factor $(N+1)^{-d}$.

\begin{figure}[t]
{\centering
\resizebox*{0.90\columnwidth}{!}{\includegraphics{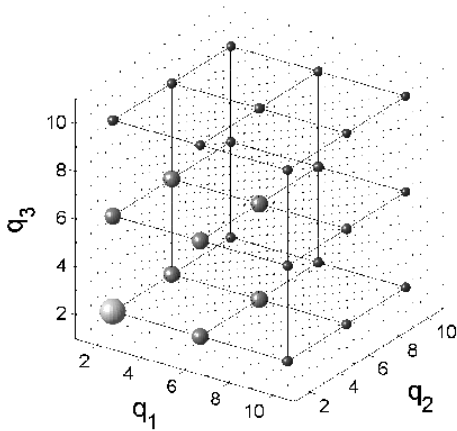}}}
{\caption{The structure of the QB with $\boldsymbol{q}_0=(2,2,2)$
on the three-dimensional lattice $N=11, \beta=0.5, E_{tot}=1.5,
\hat{\Omega}\approx0.897$. The size of spheres is a linear
function of the decimal logarithm of the linear mode energy
$E_{\boldsymbol{q}}$.}\label{fig5a}}
\end{figure}

\begin{figure}[t]
{\centering
\resizebox*{0.90\columnwidth}{!}{\includegraphics{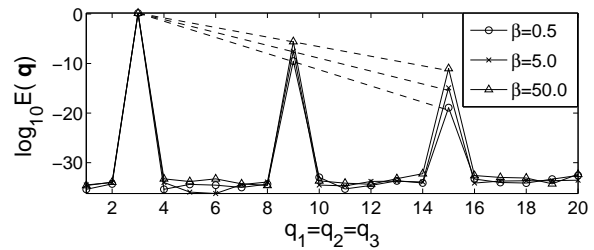}}}
{\caption{ The normal mode energies along the leading direction
$\boldsymbol{q}(n)=(n,n,n)$  for
$\boldsymbol{q}_0=(3,3,3), E_{tot}=1.5, N=20$ and the analytical
estimate (\ref{eq4}) (dashed lines).} \label{fig6}}
\end{figure}

In conclusion, we report on the existence and remarkable
properties of $q$-breathers as exact time-periodic solutions in
nonlinear two- and three-dimensional acoustic systems, thus
extending the concept of one-dimensional QBs to more physically
relevant objects. They are exponentially
localized in the $q$-space of the normal modes and preserve
stability for small enough nonlinearity. In the limit of infinite
lattice size the localization length stays constant for $d=2$ and
tends to zero for $d=3$ when fixing $\boldsymbol{q}_0$ and 
the total energy. At the same time the localization length does
not depend on the lattice size $N$ when fixing the energy density
and the wave vector $\boldsymbol{\kappa}_0=\boldsymbol{q}_0/(N+1)$. In contrast
to the one-dimensional case the observed instability threshold in
the nonlinear coupling strength increases with increasing lattice
size, and stays constant if the energy density is fixed.

M.I., O.K., and K.M. appreciate the warm hospitality of the Max
Planck Institute for the Physics of Complex Systems. M.I. and O.K.
acknowledges support of the "Dynasty" foundation.

\end{document}